\documentclass[12pt]{article}
\usepackage{graphicx,epsfig,amssymb,color}
\title{The nuclear pore complex as an
entropic gate: theory and simulation}
\author{
Mike Castellano$^a$, Steffen Wolf$^b$, Thorsten Koslowski$^{a,*}$
\\ \ \\
$^a$Institut f\"ur Physikalische Chemie, 
Universit\"at Freiburg, \\ 
Albertstra\ss e 21, 79104 Freiburg im Breisgau, Germany
\\ \ \\
$^b$Institut f\"ur Physik, 
Universit\"at Freiburg, \\ 
Hermann-Herder-Stra\ss e 3, 
79104 Freiburg im Breisgau, \\
Germany
\\ \ \\
}
\date{\ }

\begin{document}
\maketitle
*Corresponding author
\newpage
\begin{abstract}
Protein chains of the (FG)$_n$ ($n \simeq$ 300) type cap
the cytoplasmatic side of the nucleopore complex, which connects
the nucleus to the remainder of an eukaryotic cell. We study the
properties of three fundamental
polymer models that represent these
filaments using Monte Carlo computer simulations. Random walks
and the worm like chain model cannot account for the 
unusual size selectivity of the pore, while a two-dimensional 
arrangement of intrinsically disordered block copolymers with 
a high content of $\alpha$-helices is in agreement with the 
biochemical findings. We predict a linear increase of the 
free energy barrier of protein transport through the pore
with increasing protein diameter, which can be probed 
experimentally using atomic force microscopy or optical tweezers. 
\end{abstract}
% -----------------------------------------------------------------

\newpage
\subsection*{1. Introduction}

In eukaryotic cells, the genetic material is located within
the nucleus of the cell, which is separated from the 
remainder of the cell by a double membrane. This membrane is
penetrated by several thousand nuclear pore complexes (NPCs), 
which enable the export of mRNA from
the nucleus, and the import and export of proteins
\cite{Peters2005,Lim2006,Kabachinski2015,Knockenhauer2015,Lin2019}.

In figure 1, we show the structural elements of the NPC as far
as relevant to our work. The part of the pore that bridges the
double membrane of the nucleus has an outer diameter of $\sim 120$
nm, the inner diameter of the pore amounts to $\sim 52$ nm. The
total height of the complex
equals $\sim 40$ nm excluding the so-called basket
on the side of the nucleus, which is not shown here. On the
cytoplasmatic side, the NPC is capped by eight filaments
with a repetative sequence of glycine and phenylalanine 
amino acids, (FG)$_n$, with $\simeq$ 3000 iand a chain 
thus containing
roughly 600 amino acids.
The numbers given here represent the predominant variant
of the human NPC, but pore dimensions or filament 
size can differ from organism to organism \cite{Rout1993}.

NPCs exhibit a remarkable size selectivity while importing
proteins. Typically, proteins with a mass less than 4 kDa 
pass the membrane within seconds. With a mass of 17 kDa, 
the transfer may take minutes, and a mass of 40 kDa 
requires binding to auxiliary proteins -- the
appropriately named importins -- to pass the pore
\cite{Keminer1999}. Assuming a protein
density of 1.35 g/cm$^3$ \cite{Fischer2004} 
and approximating the protein
as a sphere, the molecular weights translate into diameters 
of 2.3, 3.4 and 4.5 nm, respectively. 

Solving the structure of the individual NPC proteins and  
arranging them into a global model has been one of the
major success stories of structural biology 
\cite{Eibauer2015,Kosinski2016}.
To this success, progress in experimental cryo-transmission
electron microscopy and data analysis have contributed
significantly. While we can be confident that the ring and 
the basket parts of the pore are sufficiently rigid to be 
explored by the tools of structural biology, the situation is less
clear for the FG filaments.
Typically, in polypeptides F tends to be incorported
in $\alpha$-helices, and G breaks ordered secondary structures. 
This may introduce random orientations, resulting
in chains with possible near-order structural elements,
but missing a defined global structure.
 
The repetative sequence of the FG filaments suggests to
view them through the eyes of theoretical polymer chemistry
and physics. 
The protein to be imported occupies a volume
in space that is available to the polymer chains in its 
absence. Hence, the number of 
conformations that the polymers can explore is reduced
while the protein is in reach of the chains. 
In turn, the entropy 
of the ensemble of chains is reduced, and the free energy 
of the system increases. This concept now has a strong
experimental 
angle, as nanoscopic objects can be studied and manipulated
using atomic force microscopy \cite{Ott2017} or 
optical tweezers \cite{Pang2012}, scanning the force that is
operative as a function of the distance. 

From a theoretical perspective,
atomistic molecular dynamics simulations 
can also be used to get insight into complex biochemical 
systems. They are limited in the size and in the time scale
that can be explored, and they rely on the force field
underlying the simulations. With continuous progress
in simulation algorithms and computer power, 
systems of the size of the NPC now lie within the
range of simulations. In landmark work, 
Miao and Schulten have studied the components of the 
pore \cite{Miao2009,Miao2010}. 
They have found 
the arrangement of chains into a disordered brush
repelling large objects entering the pore.
In a large-scale efford, Ando and Gopinathan
have simulated the entire yeast NPC, deriving a complex scheme
for protein transport \cite{Ando2017}.

In the work presented here, we make the attempt to
reduce the complexity of the system by inspecting 
one of its constituents, the FG filaments
using coarse-grained, strongly reduced
models usually at home in polymer chemistry
and physics.
As input, they
require a very small number of parameters, which can often
be obtained from conceptually simple experiments, such as
small angle scattering or the measurement of elastic
properties, as detailed and referenced below. 
In this way, we are able to verify or falsify
the applicability of the models and make a statement 
about the structural properties of the filaments. We 
restrict our study to blockade effects, leaving the
more
complex biochemistry of importing large proteins aside
in the simulations. We will, however, return to this 
point in ths conclusions section.

In our approach,
we take the configurational entropy as the only 
contribution to the free energy, and compute the resulting
effective interaction of an idealized protein 
-- a sphere of radius $R_{prot}$ -- with the chains.
Let $p$ be the probability of a single polymer not overlapping 
with the protein. 
We note that the probabilty of $m$ filaments not overlapping
with the protein amounts to $p^m$.
We then have
\[
\Delta G = -T\Delta S = -k_BT \ln p^m ,
\qquad (1)
\]
which forms the basis of our simulations. 
Following the 
model-specific rules detailed below, filaments are constructed 
and probed for an overlap with the protein. In turn, $p$
and $\Delta G$ are computed. In our model, the filaments do not
interact, but their impact on gating is cumulative.

% -------------------------------------------------------------

\subsection*{2. Models and methods}

% random walk
In the following, we present the polymer models used in 
this work and motivate the choice of their parameters. 
They are depicted schematically as short chains in figure 2.
As one
of the simplest polymer models, we consider the classical
ideal chain or pure random walk of $n$ monomers with an
individual length $L$. The contour length of the 
polymer is given by $\ell=nL$ (figure 2a). In two and three 
dimensions, the model gives rise to a scaling behaviour 
of the radius of gyration or the end-to-end distance as
$R_G \propto R_{ee} \propto n^{1/2}$. Experimentally, the underlying parameters
have been determined by 
small-angle x-ray scattering on a set of 33 denatured proteins
that span a large spectrum of sequence lengths
\cite{Kohn2004,Kohn2005}. With 0.598, 
the exponent differs little from ideal behaviour, and the 
average length of a monomer amounts to 1.93 \AA .
We note that the largest protein studied by
Kohn et al. \cite{Kohn2004,Kohn2005}, 
GroEL, contains 588 amino acids, which is 
close to the number of monomers in the FG filaments under
review here.

% WLC
In a second approach, we study the worm like chain (WLC) 
model in the discrete version of Kratky and Porod
\cite{Kratky1949} (figure 2b).
Here, two neighbouring monomers $i$
and $j$ experience an interaction that is
proportional to the mutual orientation of the two segments,
\[
V_{ij}=-V_0 \cos \theta_{ij}
\qquad (2)
\]
The WLC model gives rise to a squared end-to-end distance 
\[
R_{ee}^2 = 2P \ell \left[ 1 - \frac{P}{\ell} 
\left( 1- e^{-\ell/P} \right) \right]
\qquad (3)
\]
with the thus defined persistence length $P$ \cite{doi}.
A large $P$ is characteristic of
an elastic rod-like polymer, a vanishing $P$ recovers the 
ideal random walk model. $P$ is not only related to the structural,
but also to the elastic properties of the polymer. In this way,
it can be determined experimentally - as for double-stranded RNA -
or estimated on the basis of atomistic molecular dynamics 
simulations using a classical force field. For a 
protein $\alpha$-helix, Choe and Suna find 
$P \simeq 100$ nm by the molecular dynamics approach 
\cite{Choe2005},
about twice the value of the RNA
persistence length. To a large 
extent, this value is independent of the primary sequence
of the protein. For the contour length of an $\alpha$-helix, 
we have $\ell=n \times$1.43 \AA ,
which leads to $R$=75 nm for a free FG filament using
eq. 3.

A large value of $P$ implies a strong repulsive
interaction between neighbouring monomers. In this regime,
the potential eq. 2 can be 
expanded around the minimum of the potential energy at $\theta_0=\pi$,
\[
V_{ij} \simeq -V_0 
\left( 1- \frac{1}{2} (\theta_{ij}-\theta_0)^2 \right) . 
\qquad (4)
\]
This is tantamount to drawing the angle between two monomers
from a Gaussian distribution, as we have a Boltzmann probabilty
of finding an angle given as
\[
p(\theta) = \exp \left( -\frac{V_{ij}}{k_BT} \right)
\propto \exp \left( -\frac{V_0(\theta_{ij}-\theta_0)^2}{2k_BT} \right)
= 
\exp \left( -\frac{(\theta_{ij}-\theta_0)^2}{2\sigma^2} \right)
\qquad (5)
\]
with the variance of the Gaussian,
$\sigma = (k_BT/V_0)^{1/2}$. As it is straightforward to
generate a sequence of random numbers drawn from a Gaussian
distribution \cite{Box1958}, 
we follow this strategy in our simulations. The variance
$\sigma$ is calibrated to reproduce the end-to-end distance of 75 nm
resulting from the WLC model, eq. 3. As described in the 
supporting information, we arrive at $\sigma$=2.8 degrees.
This corresponds to a $V_0$ value of 
0.075 kcal mol$^{-1}$ degrees$^{-2}$.

% block copolymer, concept
As a third model, we inspect a block copolymer \cite{Jenkins1969}
which consists of both rigid helix and random walk elements 
(figure 2c). 
We consider a FG filament as an intrinsically disordered polymer
\cite{Tompa2009}, where the secondary structure
elements fluctuate with time or within a thermodynamic ensemble of
chains. 
% block copolymer, Ising model
Its energy is described using a nearest-neighbour 
Ising-like model, where the indices
$i$ represent bonds between amino acids, which either
lie within an $\alpha$-helix or within a random coil. 
We have  
\[
H = -J \sum_i S_i S_{i+1}
\qquad (6)
\]
with couplings $J$ between nearest neighbour bonds. The
$S_i$ encode the secondary structure, with $S_i$=1 for
helices and $S_i$=0 for other structural elements.
We do not consider $\beta$-sheets, as they only play
a minor role in the secondary structure, as suggested by
the Robetta structure predictions described below.
$J$ is positive, and thus the formation of helices
is favoured. Monomer lengths are 1.43 \AA \ for helices
and 1.93 \AA \ otherwise, in accord with the models 
described above. For large values of $J^*=J/k_BT$, this
model essentially becomes a defect model, where long
helices are changing their orientation at
junctions defined by the defects. 
We are aware of the more complex
nature of the chemical bond within $\alpha$-helices,
which is mediated by a strong non-covalent hydrogen
bond between 1-4 (or third-nearest) neighbours.
Nevertheless, we are confident that on the large length 
scale of some ten nanometers, a coarse-grained
model is applicable.

% Monte Carlo simulations - all three models
All of the three polymer models are simulated by 
Monte Carlo procedures. Initially, the head of the polymer chain
is placed on the circle that defines the pore, and the polymer is
build by adding monomers consecutively.
For the ideal chain, the monomer $i+1$ is randomly placed on
a circle (2d model) or sphere (3d model) of a 
radius $L$ centered at the position of the $i$'th monomer. 
For the Kratky-Porod model, a sequence of inter-monomer
angles is generated, which forms the input of a standard 
structure builder based on the TINKER molecular
modeling package. Dihedral angles are drawn from a binary
distribution (zero or 180 degrees) within the 2d model, 
or from a uniform distribution of angles in 3d.

The construction of the block copolymer model is 
performed in two steps. First, the secondary structure 
is simulated using the Ising-like model, equ. 6. 
From this simulation, statistically independent snapshots
are taken. Based on these snapshots, the geometry is 
constructed as either adding a monomer in a random 
direction ($S_i=0$), or by prolonging a rigid linear 
chain ($S_i=1)$. 

% --------------------------------------------------------------- 

\subsection*{3. Results and discussion}

In our simulations, 
all filament models have been grafted to the interior of 
the cytoplasmatic ring. 
Continuous random walks (CRWs) in two and three dimensions
have been simulated using 10$^6$ realizations, of which
at least
several thousand do not overlap with the membrane or
pore wall in two dimensions. 
In three dimensions, we typically find
10$^4$ realizations that neither overlap 
with the membrane nor collide with the
walls of the pore. Only the conformations not overlapping
with the membrane or protein wall have been considered 
for the computation of the free energy according to eq. 1.
The domains of overlap lie outside a circle in two dimensions,
and outside a cylinder and within the membrane in three dimensions.
The sphere representing the protein
is always centered in the pore. It is located at 
the entry of the cytoplasmatic side of the pore where
$p$, the probability of not overlapping with the sphere,
becomes a minimum. The geometry is illustrated in figure 3.

For random walk models of the (FG)$_n$ polypeptide, 
the computed free energy, eq. 1,
is shown in figure 4 as a function of the radius of
the sphere. It is virtually zero for a protein radius 
smaller than 14 nm , and it rises steeply at a larger 
radius. We find a barrier height equal to the thermal energy
$k_BT$ at $\sim$ 18 nm in two dimensions, which is shifted 
towards $\sim $ 21 nm  in three dimensions. The 
continuous random walks do not show any size-dependent
selectivity effects below 12 nm and hence can be 
ruled out as models of the FG filament structure. 
These findings can be easily rationalized, as free 
random walks that are described by the model parameters
used here exhibit an and-to-end distance of 
$R_{ee}=\sqrt{n}L=\sqrt{600}\times 1.43$\AA \ = 3.5 nm. 
This value cannot be expected to increase by orders of magnitude
while slightly constraining the configuration space. 
Hence, entropic repulsion by CRWs
becomes operative at a length scale 
f $R_{pore}-R_{ee} \simeq $ 20 nm.

For 2$\times$10$^6$ realizations of the 
two-dimensional Kratky-Porod model, we find a very 
small fraction of filament realizations ($\sim$ 1000) that
do not show overlap with the membrane or the wall of the pore.
This behaviour can be easily rationalized inspecting
the average end-to-end distance of the corresponding 
free chain, $R_{ee}=75$ nm, which is slightly
larger than the pore diameter of 52 nm. 
Nonetheless, the barrier for protein transport is
not very steep for this model, and we reach 
$\Delta G=k_BT$ at $R_{prot}$=15 nm . 
The situation 
is different for the three-dimensional Kratky-Porod model
(KPM), where the fraction of chains not overlapping with the
membrane or the wall of the pore is comparable
to that of the CRW models. Overlap even with large model
proteins is, however, small. The correponding
barrier is always smaller than $k_BT$ even for protein
radii very close to the pore radius, cf. fig. 4.

The properties of the two-dimensional
block copolymer model depend on
the dimensionless coupling parameter, $J^*$.
A reasonable choice for $J^*$ would be the free 
energy content of the formation of a hydrogen 
bond between two amide groups in water, which has been found 
in a range of ca. 2-8 $k_BT$ \cite{BenTal1997,Sheu2003}.
For a moderate $J^*$=1.0, the barrier 
steeply increases with an increasing protein radius.
The barrier is considerably higher than that of the 
continuous random walks and the three-dimensional 
Kratky-Porod model. It is, however, considerably smaller
than $k_BT$ in the range of radii where the size
selectivity is operative, i.e. between 1.1 
and 2.4 nm. It can be shifted into that range by
increasing $J^*$ to 2.2, corresponding to the
lower end of amide hydrogen bond free energies.
Here, we find
a barrier that is with increasing protein radius, and we 
have $\Delta G = k_BT$ at $R_{prot}\simeq$ 1 nm,
a barrier that can be easily passed thermally.

Relative errors of the simulation methods have been evaluated 
at free energy barriers of $k_BT$ by simulating 50 realizations
using the same parameters and number of Monte Carlo
steps as in the production runs. Via the root mean square 
variations, we find relative errors of 0.04 (continuous
random walk, 2d), 0.03 (continuous random walk, 3d), 
0.08 (Kratky-Porod model in 2d, 
block copolymer with $J^*$=2.2) and
0.11 (Kratky-Porod model in 2d, 
block copolymer with $J^*$=1.0).
For the three-dimensional Kratky-Porod model, this 
quantity has been computed as 0.03 at a barrier heigth
of $k_BT/2$.

In addition to the Monte Carlo simulations, we have inspected
(FG)$_{20}$ oligomers from a bioinformatics angle. We have used
the Robetta suite, which combines homology modelling and a 
{\sl de novo} fragment insertion method \cite{Kim2004}. 
The results of five
predicted structural models are presented in figure 5 
and in table 1. The assignment of the secondary structure elements
has been made with the help of the 2Struc program 
\cite{Klose2010}. This analysis is largely heuristic and is
to viewed with some caution. Nevertheless, it provides additional
information on the structure of the FG filaments. All but one model 
predict a mixture of $\alpha$-helical and random 
secondary structure elements, with the exception of one model
that finds a small contribution of $\beta$-sheets.

% --------------------------------------------------------------- 

\subsection*{4. Conclusions}

To get insight into the nature of the gating mechanism 
of the nuclear pore complex, we have inspected simple
models of polymer science to describe the FG filaments
capping the complex. Their space of conformations
is restricted by a spherical model protein passing
the pore, giving rise to an entropic barrier. 
The systems have been simulated by Monte Carlo methods.
The models are checked 
qualitatively against the
peculiar size selectivity of the pore. The
selectivity  sets in at
protein diameters in the range of 3-5 nm, values that 
are considerably smaller than the pore diameter of 52 nm. 
In the range of interest, continuous random walk
models and the Kratky-Porod model lead to barriers that are
much smaller than $k_BT$ and can hence be easily overcome
by diffusion. This statement holds both for two- and
three-dimensional variants of the models. 

On the other hand, a two-dimensional block copolymer 
model of the filaments
shows promising features interacting with comparatively
small proteins. It predominatly consists of rigid
$\alpha$-helices that are linked
by disordered structural elements. In the model, the 
disorder induces a reorientation of the protein, 
cf. figure 2. In the parameter range of interest, it is
reduced to a handfull of defects 
($\sim$ 1-2 \% ) that break the helix
and its orientation. With increasing
protein diameter, the free energy barrier increases 
linearily. 
It equals $k_BT$ at a small 
length scale of $R_{prot}\sim$ 0.75  nm.
This does not rule out the applicability of polymer
models not tested. In particular, in two dimensions
different models may show a very similar scaling behaviour
\cite{Hsu2011}.

The structural elements of block copolymer
model also dominate
homology models of short FG oligomers, albeit with
a different weight. Experimental structures of the 
proteins mediating protein transport through the 
pores, the importins, are available
\cite{Kobe1999,Cingolani1999}. It is interesting
to note they also contain FG repeats with a
high content of $\alpha$-helices. Given our findings,
it might be the role of importins to locally stabilize
defects in the block copolymer, thus locally lowering 
$J^*$ and the energetical barrier for diffusion through 
the NPC. 

Our interpretation of the block copolymers is a
dynamic one. The defect positions change with time,
and in a thermodynamic ensemble many different
realizations will coexist. From this perspective, the 
FG filaments can also be viewed as intrinsically
disordered proteins. In the limit of a defect model,
the probabilty distribution of finding 
a helix of a certain length among the amino acids
of the filament is a long-tailed one, cf. the supporting
information. Under these
conditions, a random walk consisting of large
rigid helices explores space much more
efficiently than a standard random walk with equal step
size. For the FG filaments, this is tantamount to a very
efficient blockade of the pore. 
Such a process has been referred to 
as a Levy flight by Mandelbrot \cite{Mandelbrot1982}.

From our perspective, the gating function of the nucleopore 
FG filaments provides a rare example in protein biochemistry:
here, the function of a protein is not based on the 
specific chemistry of an elaborate
sequence of amino acids fine-tuned by evolution, but on a 
simple, repetative pattern that mainly works according to the laws
of polymer physics.

% --------------------------------------------------------------- 

\subsubsection*{Acknowledgements}

It is a pleasure to thank T. Friedrich, T. Hugel, K. Jahnke, 
J. Kaiser and N. Kremer for fruitful discussions

\subsubsection*{Data Availability Statement}

The data that support the findings of this study are 
available from the corresponding author upon reasonable request.

% --------------------------------------------------------------- 
\newpage
\subsubsection*{Table 1}

Secondary structure analysis of homology models 
of (FG)$_{20}$ oligomers, as predicted by the 
Robetta suite \cite{Kim2004}.
       
\begin{tabular}{l|r|r|r|r|r}
model &  1    & 2    &  3    &   4   &  5  \\
\hline
helix &  63 & 20 &  26 &  34 & 57  \\
sheet &   0 & 14 &   0 &   0 &  0  \\
other &  37 & 66 &  74 &  66 & 43  \\

\end{tabular}

% --------------------------------------------------

\newpage

\subsection*{Figure captions}

{\bf Figure 1}

Model of the nuclear pore complex showing the 
constituents relevant to our work. 
a. FG filaments,
b. cytoplasmatic ring,
c. central framework,
d. nuclear ring,
e. top view.

{\bf Figure 2}

Short sequences of
polymer models used in this work. 
a. continuous random walk,
b. the Kratky-Porod model as the 
discrete version of the worm like chain model and
c. block copolymer model.

{\bf Figure 3}

Illustration of acceptance and rejection within the 
two-dimensional Monte Carlo simulations. a. membrane and
nuclear pore complex wall, b. pore,
c. protein, d. accepted polymer conformation, e. rejected 
polymer conformation (collision with protein), f. polymer
conformation discarded due to collision with the membrane
or wall of the nuclear pore complex.

{\bf Figure 4}

Free energy barriers for the transfer of hard spheres
as a function of their radius. The symbols correspond to 
the following polymer models: 
block copolymer, 2d, $J^*$=2.2 ($\circ$),
block copolymer, 2d, $J^*$=1.0 ($\nabla$),
block copolymer, 3d, $J^*$=2.2 ($\triangle$),
continuous random walk, 2d ($+$),
continuous random walk, 3d ($\times$),
Kratky-Porod model, 2d ($\bullet$)
and the Kratky-Porod model in 3d ($\square$).

{\bf Figure 5}

Five models of secondary structure for (FG)$_{18}$, as suggested by
the Robetta program package \cite{Kim2004}
and analyzed using the DSSP approach \cite{Klose2010}.
In the color version of the image, $\alpha$-helices are drawn in red,
$\beta$-sheets are colored yellow, and others are depicted 
in green.

% --------------------------------------------------------------- 
\newpage
{\bf Figure 1}

\begin{figure}[ht]
\includegraphics[width=1.0\textwidth]{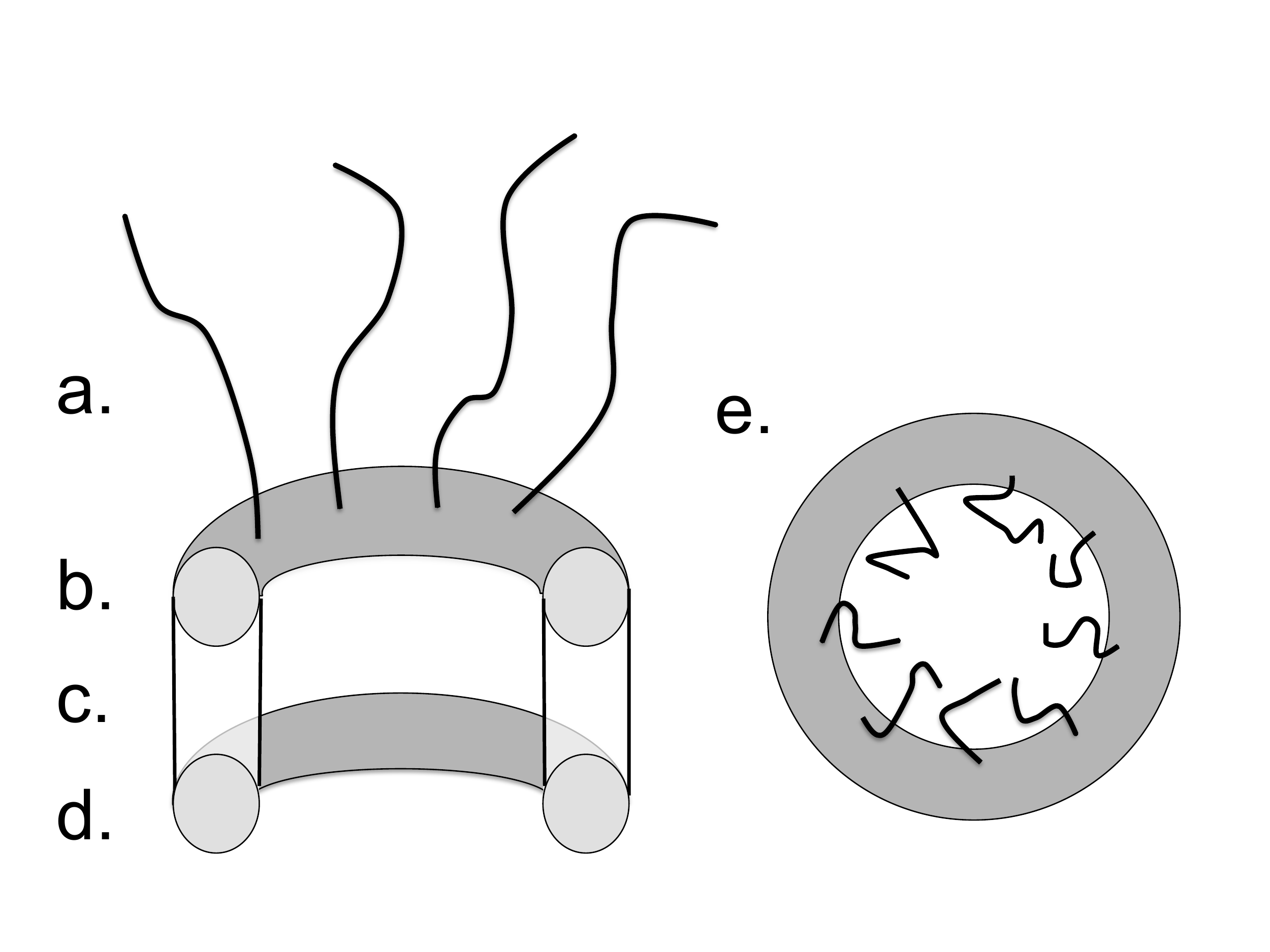}
\end{figure}
% --------------------------------------------------------------- 
\newpage
{\bf Figure 2}

\begin{figure}[ht]
\includegraphics[width=1.0\textwidth]{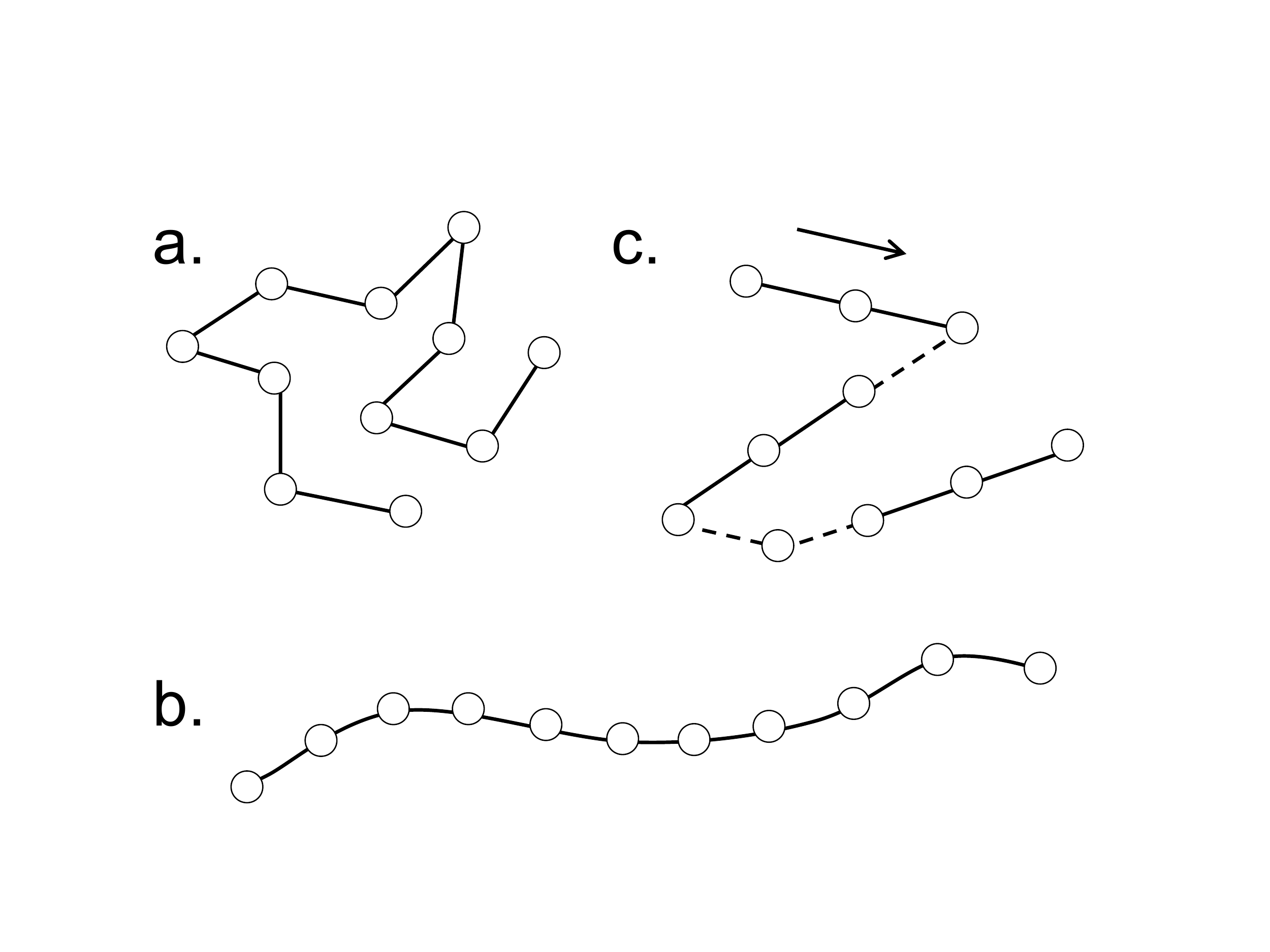}
\end{figure}
% --------------------------------------------------------------- 
\newpage
{\bf Figure 3}

\begin{figure}[ht]
\includegraphics[width=1.0\textwidth]{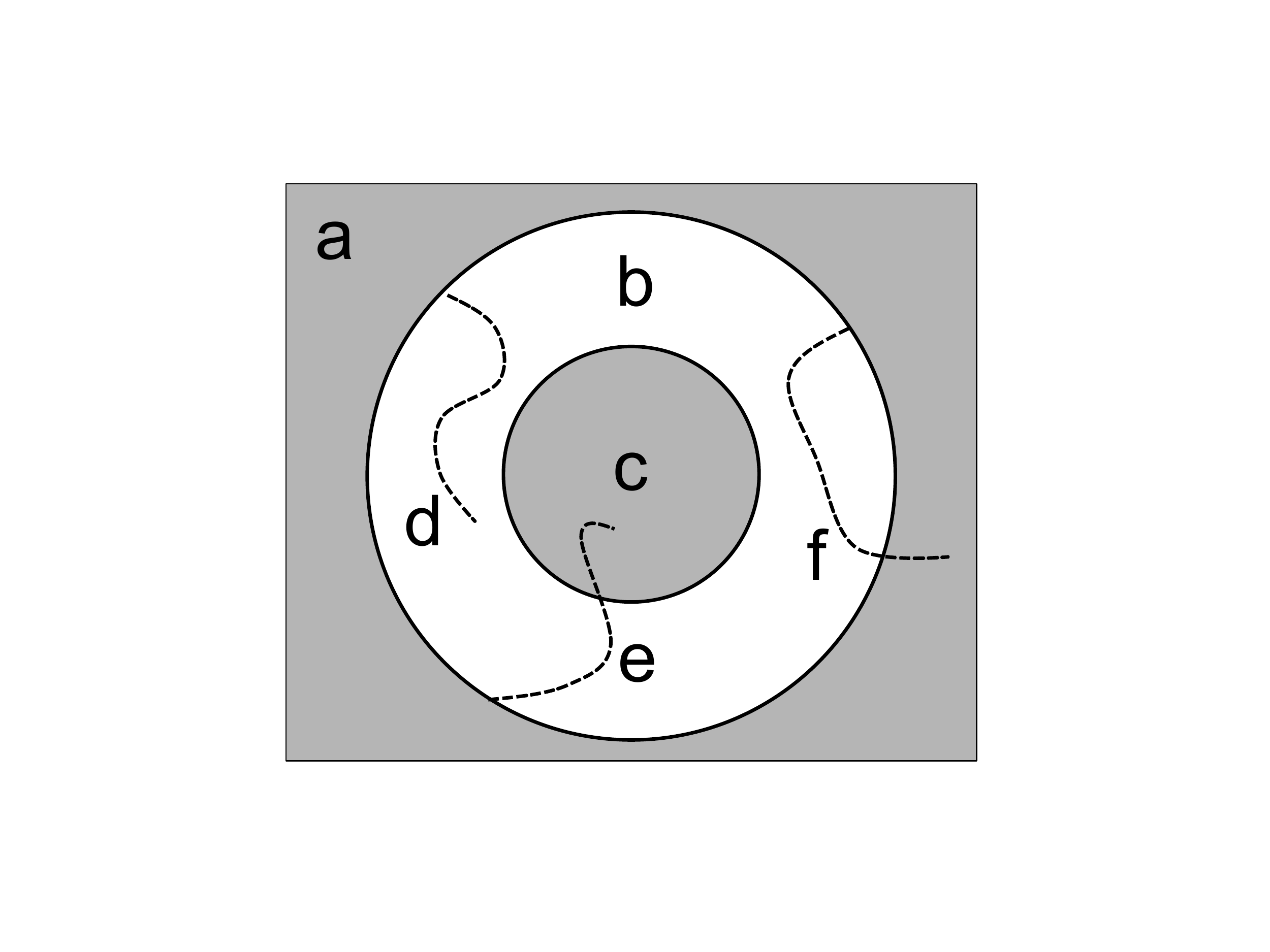}
\end{figure}
% --------------------------------------------------------------- 
\newpage
{\bf Figure 4}

\begin{figure}[ht]
\includegraphics[width=1.0\textwidth]{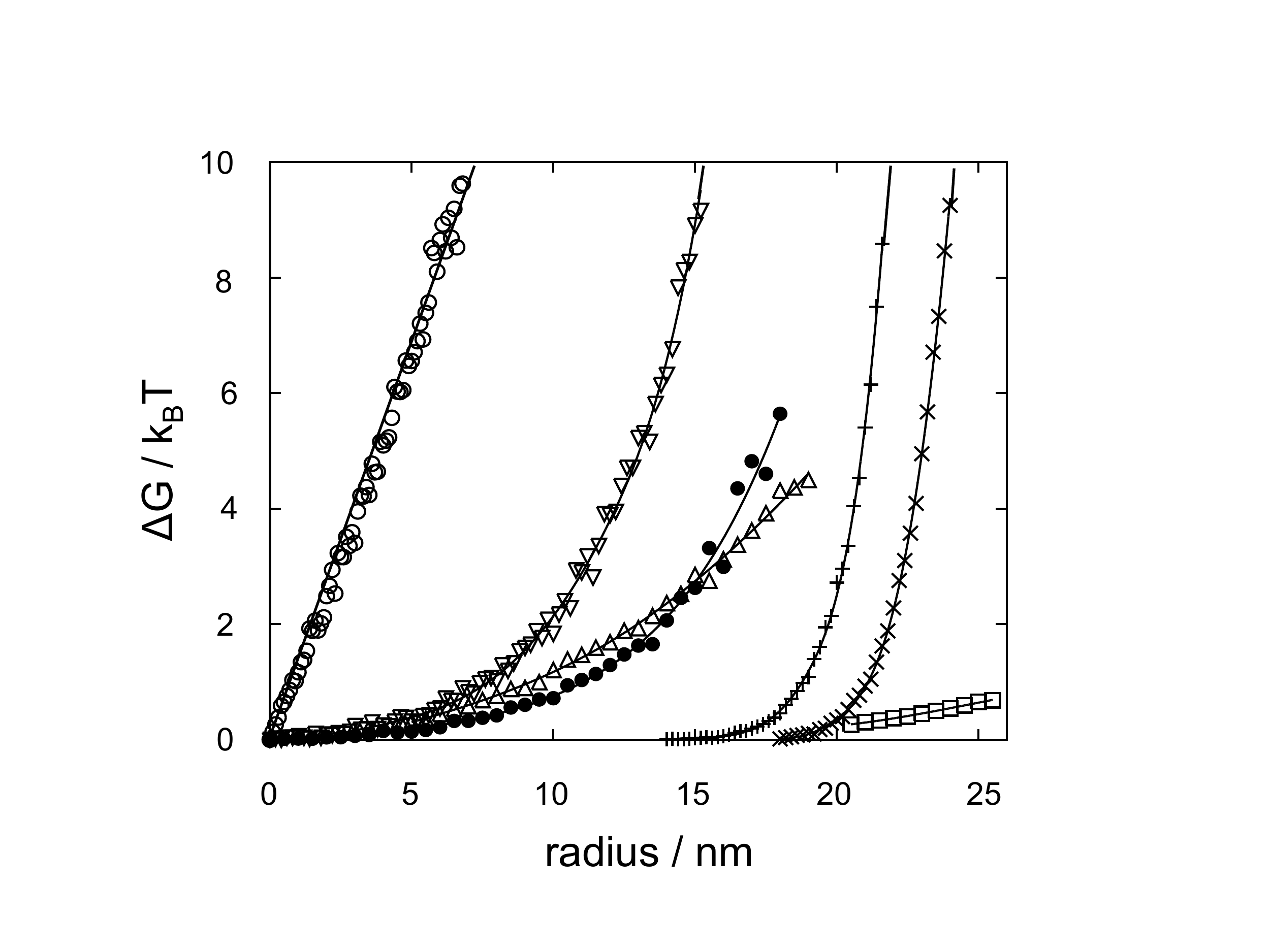}
\end{figure}
% --------------------------------------------------------------- 
\newpage
{\bf Figure 5}

\begin{figure}[ht]
\includegraphics[width=1.0\textwidth]{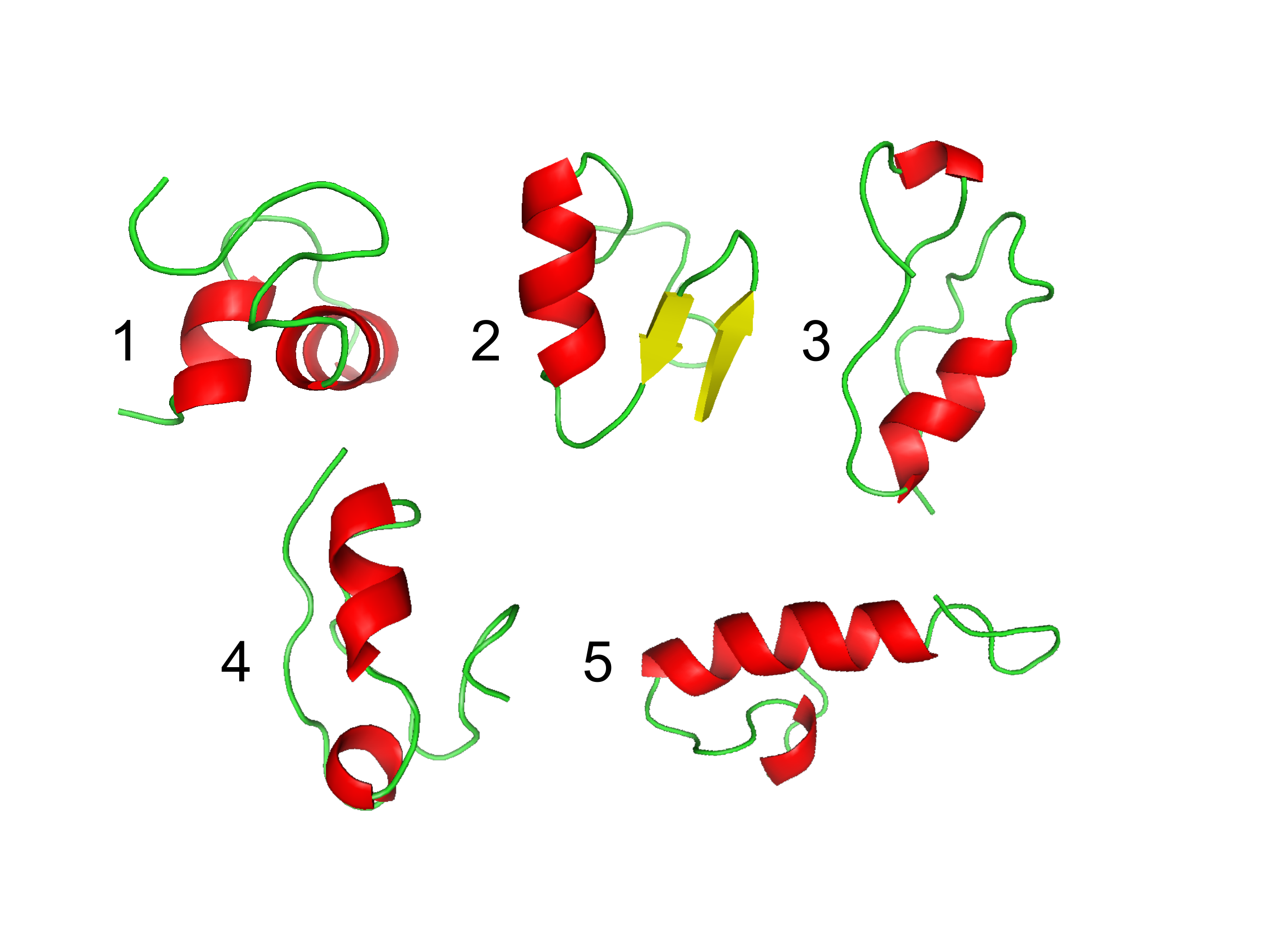}
\end{figure}
% --------------------------------------------------------------- 

\newpage
\bibliography{npcbib}{}

\begin{thebibliography}{10}

\bibitem{Peters2005}
R.~Peters.
\newblock Translocation through the nuclear pore complex: Selectivity and speed
  by reduction-of-dimensionality.
\newblock {\em Traffic}, 6:421--427, 2005.

\bibitem{Lim2006}
R.~Lim, U.~Aebi, and D.~Stoffler.
\newblock From the trap to the basket: Getting to the bottom of the nuclear
  pore complex.
\newblock {\em Chromosoma}, 115:15--26, 2006.

\bibitem{Kabachinski2015}
G.~Kabachinski and T.U. Schwartz.
\newblock The nuclear pore complex structure and function at a glance.
\newblock {\em Journal of Cell Science}, 128:423--429, 2015.

\bibitem{Knockenhauer2015}
K.E. Knockenhauer and T.U. Schwartz.
\newblock The nuclear pore complex as a flexible and dynamic gate.
\newblock {\em Cell}, 164:0092--8674, 2016.

\bibitem{Lin2019}
D.~H. Lin and A.~Hoelz.
\newblock The structure of the nuclear pore complex (an update).
\newblock {\em Annual review of biochemistry}, 88:725–783, 2019.

\bibitem{Rout1993}
M.~P. Rout and G.~Blobel.
\newblock {Isolation of the yeast nuclear pore complex.}
\newblock {\em Journal of Cell Biology}, 123:771--783, 1993.

\bibitem{Keminer1999}
O.~Keminer and R.~Peters.
\newblock Permeability of single nuclear pores.
\newblock {\em Biophys J.}, 77:217–228, 1999.

\bibitem{Fischer2004}
H.~Fischer, I.~Polikarpov, and A.F. Craievich.
\newblock Average protein density is a molecular-weight-dependent function.
\newblock {\em Protein Sci.}, 13:2825–2828, 2004.

\bibitem{Eibauer2015}
M.~Eibauer, M.~Pellanda, Y.~Turgay, A.~Dubrovsky, A.~Wild, and O.~Medalia.
\newblock Structure and gating of the nuclear pore complex.
\newblock {\em Nature Communications}, 6:2041--1723, 2015.

\bibitem{Kosinski2016}
J.~Kosinski, S.~Mosalaganti, A.~von Appen, R.~Teimer, A.L. DiGuilio, W.~Wan,
  K.H. Bui, W.J.H. Hagen, J.A.G. Briggs, J.S. Glavy, E.~Hurt, and M.~Beck.
\newblock Molecular architecture of the inner ring scaffold of the human
  nuclear pore complex.
\newblock {\em Science}, 352:363--365, 2016.

\bibitem{Ott2017}
E.~Ott, M.A. Jobst, C.~Schoeler, H.E. Gaub, and M.A. Nash.
\newblock Single-molecule force spectroscopy on polyproteins and
  receptor–ligand complexes: The current toolbox.
\newblock {\em Journal of Structural Biology}, 197:3 -- 12, 2017.

\bibitem{Pang2012}
Y.~Pang and R.~Gordon.
\newblock Optical trapping of a single protein.
\newblock {\em Nano Letters}, 12:402--406, 2012.

\bibitem{Miao2009}
L.~Miao and K.~Schulten.
\newblock Transport-related structures and processes of the nuclear pore
  complex studied through molecular dynamics.
\newblock {\em Structure}, 17:449–459, 2009.

\bibitem{Miao2010}
L.~Miao and K.~Schulten.
\newblock Probing a structural model of the nuclear pore complex channel
  through molecular dynamics.
\newblock {\em Biophysical Journal}, 98:1658 -- 1667, 2010.

\bibitem{Ando2017}
D.~Ando and A.~Gopinathan.
\newblock Cooperative interactions between different classes of disordered
  proteins play a functional role in the nuclear pore complex of baker’s
  yeast.
\newblock {\em PLOS ONE}, 12:e0169455, 2017.

\bibitem{Kohn2004}
J.E. Kohn, I.S. Millett, J.~Jacob, B.~Zagrovic, T.~M. Dillon, N.~Cingel, R.~S.
  Dothager, S.~Seifert, P.~Thiyagarajan, T.~R. Sosnick, M.~Z. Hasan, V.~S.
  Pande, I.~Ruczinski, S.~Doniach, and K.~W. Plaxco.
\newblock Random-coil behavior and the dimensions of chemically unfolded
  proteins.
\newblock {\em Proceedings of the National Academy of Sciences},
  101:12491--12496, 2004.

\bibitem{Kohn2005}
J.E. Kohn, I.S. Millett, J.~Jacob, B.~Zagrovic, T.~M. Dillon, N.~Cingel, R.~S.
  Dothager, S.~Seifert, P.~Thiyagarajan, T.~R. Sosnick, M.~Z. Hasan, V.~S.
  Pande, I.~Ruczinski, S.~Doniach, and K.~W. Plaxco.
\newblock Correction for kohn et al., random-coil behavior and the dimensions
  of chemically unfolded proteins, pnas 2004 101:12491-12496.
\newblock {\em Proceedings of the National Academy of Sciences},
  102:14475--14475, 2005.

\bibitem{Kratky1949}
O.~Kratky and G.~Porod.
\newblock R{\"o}ntgenuntersuchung gel{\"o}ster fadenmolek{\"u}le.
\newblock {\em Rec. Trav. Chim. Pays-Bas.}, 68:1106–1123, 1949.

\bibitem{doi}
M.~Doi and S.F. Edwards.
\newblock {\em The Theory of Polymer Dynamics}.
\newblock Clarendon Press, New York, 1986.

\bibitem{Choe2005}
S.~Choe and S.X. Suna.
\newblock The elasticity of $\alpha$-helices.
\newblock {\em J. Chem. Phys.}, 122:244912, 2005.

\bibitem{Box1958}
G.E.P. Box and M.E. Muller.
\newblock A note on the generation of random normal deviates.
\newblock {\em Ann. math. stat.}, 29:610--611, 1958.

\bibitem{Jenkins1969}
A.D. Jenkins, P.~Kratochvil, R.F.T. Stepto, and U.W. Suter.
\newblock Glossary of basic terms in polymer science (iupac recommendations
  1996).
\newblock {\em Pure and Applied Chemistry}, 68:2287--2311, 1996.

\bibitem{Tompa2009}
P.~Tompa and A.~Fersht.
\newblock {\em Structure and Function of Intrinsically Disordered Proteins}.
\newblock CRC Press, Boca Raton FL, 2009.

\bibitem{BenTal1997}
N~BenTal, D~Sitkoff, I~A Topol, A~S Yang, S~K Burt, and B~Honig.
\newblock {Free energy of amide hydrogen bond formation in vacuum, in water,
  and in liquid alkane solution}.
\newblock {\em Journal of Physical Chemistry B}, 101(3):450--457, 1997.

\bibitem{Sheu2003}
Sheh-Yi Sheu, Dah-Yen Yang, H~L Selzle, and E~W Schlag.
\newblock {Energetics of hydrogen bonds in peptides}.
\newblock {\em Proceedings of the National Academy of Sciences of the United
  States of America}, 100(22):12683--12687, 2003.

\bibitem{Kim2004}
D.~E. Kim, D.~Chivian, and D.~Baker.
\newblock Protein structure prediction and analysis using the robetta server.
\newblock {\em Nucleic Acids Research}, 32:W526--W531, 2004.

\bibitem{Klose2010}
D.~P. Klose, B.~A. Wallace, and R.~W. Janes.
\newblock 2struc: the secondary structure server.
\newblock {\em Bioinformatics}, 26:2624--2625, 2010.

\bibitem{Hsu2011}
H.-P. Hsu and P.K. Binder.
\newblock Breakdown of the kratky-porod wormlike chain model for semiflexible
  polymers in two dimensions.
\newblock {\em EPL}, 95:1--5, 2011.

\bibitem{Kobe1999}
B.~Kobe.
\newblock Autoinhibition by an internal nuclear localization signal revealed by
  the crystal structure of mammalian importin $\alpha$.
\newblock {\em Nature Structural Biology}, 6:388--397, 1999.

\bibitem{Cingolani1999}
G.~Cingolani, C.~Petosa, K.~Weis, and C.W. M{\"u}ller.
\newblock Structure of importin-$\beta$ bound to the ibb domain of
  importin-$\alpha$.
\newblock {\em Nature}, 399:221--229, 1999.

\bibitem{Mandelbrot1982}
B.M. Mandelbrot.
\newblock {\em The fractal geometry of nature}.
\newblock W.H. Freeman, 1982.

\end{thebibliography}
\bibliographystyle{unsrt}
% --------------------------------------------------------------- 
\end{document}